\documentstyle[epsfig]{mn}

\begin{document}

\newcommand{\ba}{$(b/a)^2_{\rm bar}$}
\newcommand{\basq}{$(b/a)^2_{\rm bar}$}
\newcommand{\ii}{$I_{814W}$}
\newcommand{\ib}{$I_{814W}$-band}

\title[Barred Spiral Structure in the Hubble Deep Fields]{The
Evolution of Barred Spiral Galaxies in the Hubble Deep Fields North
and South}

\author[Abraham et al.]{
  R. G. Abraham$^{1}$, 
  M. R. Merrifield$^{2,3}$, 
  R. S. Ellis$^1$,
  N. R. Tanvir$^1$ and
  J. Brinchmann$^1$\\
  $^1$Institute of Astronomy, University of Cambridge, Madingley Road, 
  Cambridge CB3 OHA\\
  $^2$School of Physics \& Astronomy, University of Nottingham,
  Nottingham NG7 2RD\\
  $^3$Department of Physics and Astronomy, University of Southampton, SO17 1BJ
} 

\date{Received:\ \ \ Accepted: }

\pagerange{\pageref{firstpage}--\pageref{lastpage}}%

\maketitle

\label{firstpage}

\begin{abstract}

The frequency of barred spiral galaxies as a function of redshift
contains important information on the gravitational influence of
stellar disks in their dark matter halos and also may distinguish
between contemporary theories for the origin of galactic bulges.  In
this paper we present a new quantitative method for determining the
strength of barred spiral structure, and verify its robustness to
redshift-dependent effects. By combining galaxy samples from the
Hubble Deep Field North with newly available data from the Hubble Deep
Field South, we are able to define a statistical sample of 18
objectively-defined low-inclination barred spiral systems with
$I_{814W}<23.2$ mag. Analysing the proportion of barred spiral
galaxies seen as a function of redshift, we find a significant decline
in the barred fraction beyond redshifts $z\simeq0.5$. The physical
significance of this effect remains unclear, but several possibilities
include dynamically hotter (or increasingly dark-matter dominated)
high-redshift discs, or an enhanced efficiency in bar destruction at
high redshifts. By investigating the formation of the ``orthogonal''
axis of Hubble's classification tuning fork, our result complements
studies of evolution in the early--late sequence, and pushes to later
epochs the redshift at which the Hubble classification sequence is
observed to be in place.

\end{abstract}


\section{INTRODUCTION}

Deep exposures with the {\em Hubble Space Telescope} (HST) and, in
particular, those undertaken through the {\em Hubble Deep Field} (HDF)
campaigns (Williams et al 1996), have enabled the study of the
morphological evolution of galaxies as a function of redshift (see
Ellis 1998 and many references therein). An important goal of such
studies is an understanding of the processes by which high redshift
systems assemble and become transformed into the population
categorised by Hubble's ``tuning fork''. As the deepest exposures
reach epochs where the Hubble classification system ceases to provide
a useful description of the galaxy population (Abraham et al 1996a),
it seems likely that the origin of morphological structure can be
understood from detailed analyses of such HST data.

Most morphological studies of faint galaxies have concentrated on
addressing the statistics of regular and irregular systems (Glazebrook
et al 1995; Abraham et al 1996a,b; Driver et al 1995,1998; Odewahn et
al 1996; Schade et al. 1995; Brinchmann et al 1998; Marleau \& Simard
1998), either probing for evolution along the early--late sequence of
the tuning fork, or discussing the role of objects best described as
altogether outside of the conventional framework of the Hubble
sequence.  Key questions include the assembly rate of field
ellipticals and the evolutionary history of spiral disks.  Only
recently, through the added signal-to-noise and multi-colour data
uniquely available in the HDF exposures, have attempts been made to
address possible evolution in {\it internal} structures of distant
galaxies. Such methods are highly appropriate ways of attacking the
same questions previously addressed from morphological studies based
on the integrated properties of faint galaxies.  For example, Abraham
et al (1998) examined the homogeneity of internal colours within HDF
field ellipticals of known spectroscopic redshift and compared the
derived star formation history for field spheroidals with those
similarly determined for galactic bulges seen in distant HDF
spirals. Provided the various selection effects can be accounted for,
such comparisons directly address the question of the order of
formation of bulges and stellar disks.

This paper is concerned with extending our earlier HDF study by
examining the role that stellar bars play in the evolutionary history
of spiral galaxies. Our analysis is motivated by the possibility of an
intimate connection between the formation of bars and the growth of
galactic bulges. Bar formation can be understood from numerical
simulations of self-gravitating disks of stars on circular orbits
which are unstable to collapse along one axis (e.g.\ Miller,
Prendergast \& Quirk 1970). More recently however, it was discovered
that bars are, themselves, unstable: they ultimately buckle
perpendicularly to the plane of the disk creating a central spheroidal
component similar to a galactic bulge (Combes et al.\ 1990, Raha et
al.\ 1991). This suggests bulges might form by secular evolution, i.e.
through the growth and subsequent collapse of bar instabilities in
cold rotating discs. This contrasts with the hitherto established view
(Eggen, Lynden-Bell, \& Sandage 1962) that galactic bulges form
through early dissipationless collapse with discs subsequently growing
via gas accretion. An absence of barred spirals at high redshifts
would pose serious challenges to secular models.

In addition to understanding the history of bulge formation discussed
above, the presence of a bar is an important indicator that the host
galaxy has sufficient disk mass (relative to that in the dark halo) for
its self-gravity to be important and that the disk material is in
well-ordered circular orbits. Accordingly, the frequency of barred
galaxies as a function of redshift has important consequences for the
proportion of galactic mass contained in dark matter halos as a
function of redshift, particularly when combined with kinematic
data (Quillen \& Sarajedini 1998; Quillen 1998).

Measuring the proportion of barred systems as a function of redshift
poses a number of challenges. Simple quantitative measures of bulk
galactic structure, such as central concentration or global asymmetry,
are adequate for placing galaxies within a one-dimensional early--late
classification sequence, but objective measures of finer details of
spiral structure are much harder to define and test. Possible
evolution in the barred population was claimed on the basis of a
visual inspection of galaxies in the northern field of the HDF
(hereafter HDF-N) by van den Bergh et al (1996).  This observation
remains controversial, since bars will, most likely, be harder to
detect at rest-frame ultraviolet wavelengths sampled in high redshift
($z > 1$) spirals. van den Bergh et al (1996) considered galaxies at
all inclinations to the very limits ($I_{814W}\sim 25$ mag) to which
visual morphological classification can be used to classify galaxies
into simple early/spiral/peculiar bins in the HDF-N. But at
$I_{814W}\sim 25$ mag photometric redshifts studies (eg. Sawicki, Lin,
\& Yee 1997; Wang et al 1998; Hogg et al 1998) indicate serious
contamination by very high-redshift spirals, and it is likely that
bars are undetectable due to both bandshifting of the rest frame and
low signal-to-noise.  Although such effects can be calibrated by
simulations, without deep images probing rest-frame optical
wavelengths at high redshifts, the validity of such simulations is
difficult to assess. A more robust test would restrict samples to
low-inclination spirals with redshifts less than $z \sim 1$, where
bandshifting effects are negligible and where signal-to-noise levels
are high enough for unambiguous bar detection.  Until the release of
the southern Hubble Deep Field (HDF-S), no adequately large deep
imaging sample has been available.

A final complication is that the fraction of {\em local} barred
spirals remains controversial. There is a continuum in apparent bar
strength in galaxies, and the strength required to merit
classification as a barred galaxy is highly subjective.  For example,
there is reasonable agreement in local catalogues that the proportion
of {\em strongly} barred galaxies is 25\%--35\%, based on the numbers
given in the {\em Revised Shapley-Ames Catalogue} (RSA; Sandage \&
Tammann 1987), the {\em Third Reference Catalogue} (RC3; de
Vaucouleurs et al. 1991), and the {\em Uppsala General Catalogue}
(UGC; Nilson 1973). However an additional 30\% of spirals are
classed as {\em weakly} barred in the RC3, substantially higher than
in the UGC or RSA.  Quite apart from taxonomical differences in
accounting for weakly barred systems, morphological classications at low
redshift have largely been based on subjective inspections of
photographic data with limited dynamic range, leading to poor
agreement in the classifications made by observers using the same
classification system (Naim et al. 1995).

In the present paper we demonstrate how these difficulties can be
surmounted through quantitative measures of bar strength for a large
sample of spiral galaxies culled from both Hubble Deep Fields.  It is
clear that since the proportion of locally barred spirals is poorly
defined, the optimal way to probe for evolution in the bar fraction is
an objective study that encompasses internally a broad range of
redshifts.  The recent release of the HDF-S has doubled the size of a
suitably deep sample and has motivated the current analysis.  A plan
of the paper follows. In Section~2 we describe our objective measure
of bar strength, outline its physical significance, and show how this
bar-strength measure has been calibrated using local galaxy samples.
In Section~3 we demonstrate the robustness of our bar strength
parameter, and calculate the limits to which strongly barred systems
should be detectable in the Hubble Deep Fields.  The methodology from
Abraham et al (1996) is used to select from the HDF samples a subset
of low-inclination spirals whose images have sufficient
signal-to-noise to allow barred structure to be detected.  We then
determine the fraction of barred galaxies in the Northern and Southern
Hubble Deep Fields as a function of spectroscopic and photometric
redshift, and conclude that a strong evolutionary effect exists. In
Section~4 we discuss the implications (and associated uncertainties)
of our results for models of the formation of bars, bulges, and disks
at high redshifts. Our conclusions are summarised in Section~5.
 
\section{A QUANTITATIVE MEASURE OF BAR STRENGTH}

\subsection{Methodology}

Characterising spiral structure in distant systems poses a challenge
because of the diversity of arm structures that are seen locally, and
because spiral arms can be difficult to observe at low signal-to-noise
levels where surface brightness dimming becomes important.  However,
in the absence of strong bandshifting effects (the importance of which
are discussed in \S2.2), galactic bars should be the easiest of the
spiral features to detect in high resolution data because of their inherent
brightness and symmetry about the central nucleus.

The presence of a bar in a galaxy will have two measurable effects on
its photometry: generally speaking, the ellipticity of the isophotes
will change between the region of the bar and the outer galaxy, and
the principal axes of the isophotes will also vary between these two
regions.  In order to quantify the strength of these signatures in
distant galaxies, we have adopted the following procedure.  Galaxy
images are first isolated from the sky background by extracting
contiguous pixels at $1.5\sigma$ above the sky level.  Each galaxy
image is then ``sliced'' at 1\% and 85\% of its maximum flux
level. The second-order moments of the pixels brighter than these flux
thresholds define two best-fitting ellipses, the first for the entire
galaxy and the second for the inner portion of the galaxy near the
nucleus.  From the parameters of these best-fit ellipses, we extract
the axis ratio of the galaxy as a whole, $(b/a)_{\rm outer}$, and that
of the inner part of the galaxy, $(b/a)_{\rm inner}$.  We also measure
the twist angle between the principle axes of these ellipses,
$\phi=\phi_{85\%}$ ($0 < \phi < 90^{\circ}$).

In the vast majority of cases, the cut at 85\% of the maximum flux
proves an effective level for isolating any bar-like structures.
However, in a few obviously-barred galaxies with bright nuclei (often
systems with dominant rings, ``fat'' bars, or morphological lenses)
this cut lies inside the strongly-barred region.  We therefore repeat
the analysis with the inner cut at a flux level of 50\% of the maximum.
If this choice of flux level were to detect a bar that the original
choice missed, then we would expect the twist between the ellipses in
this analysis, $\phi_{50\%}$, to be much greater than $\phi_{85\%}$.
Therefore, if $\phi_{50\%} > 10 \times \phi_{85\%}$, then we adopt this
lower flux cut to determine $(b/a)_{\rm inner}$, and define the
twist angle by $\phi=\phi_{50\%}$.

We now need some objective criterion for interpreting the measured
values of $(b/a)_{\rm inner}$, $(b/a)_{\rm outer}$ and $\phi$ as a
measure of the ``barriness'' of a galaxy.  To do so, we have chosen to
measure a parameter related to {\em the physical axial ratio of the
bar} in an idealized galaxy.  Consider a simple model for a galaxy in
which at large radii it is an axisymmetric thin disk, so that its
inclination, $i$, is directly related to $(b/a)_{\rm outer}$:
\begin{equation}
i = \cos^{-1} \bigl[(b/a)_{\rm outer}\bigr].
\end{equation}
\noindent
If we further assume that the inner region quantified by $(b/a)_{\rm
inner}$ can be modelled by an elliptical distribution of light lying
in the same plane as the outer disk, then we can infer its intrinsic
axis ratio, $(b/a)_{\rm bar}$, from the measured values of $(b/a)_{\rm
inner}$, $\phi$, and $i$.  After the appropriate coordinate
transformations have been made to rotate the galaxy to face-on, we
find that
\begin{equation}
(b/a)_{\rm bar}^2 = {1 \over 2}\left(X - \sqrt{X^2 - 4}\right),
\end{equation}
\noindent where
\begin{eqnarray}
X & = & \sec^2i \bigl[ 2\cos^2 \phi \sin^2 \phi \sin^4i \nonumber\\
  &   & \quad\qquad + (b/a)_{\rm inner}^2(1 - \sin^2 \phi \sin^2i)^2\nonumber\\
  &   & \quad\qquad + (b/a)_{\rm inner}^{-2}(1 - \cos^2 \phi \sin^2i)^2\bigr].
\end{eqnarray}
				  
Formally, these formulae only yield the axis ratio for a bar in the
idealized case where it is flat, exactly elliptical, and embedded in a
circular disk.  However, as we will show in the next section, the
value of $(b/a)_{\rm bar}^2$ still provides a robust, objective
measure of ``barriness'' when these ideal conditions are relaxed: real
galaxies that the eye recognizes as barred have systematically larger
values of $(b/a)_{\rm bar}^2$ than do unbarred systems.  There is,
however, one important caveat to the general applicability of this
formula: the assumption of a two-dimensional disk clearly becomes
unreasonable if a galaxy is viewed close to edge-on, when the
three-dimensional shape of the central bulge becomes a prominent
feature.  However, it is intrinsically almost impossible to determine
whether a galaxy that lies close to edge-on is barred on the basis of
photometry alone.  We therefore only attempt to determine the value of
$(b/a)_{\rm bar}^2$ for galaxies where we derive an inclination of $i
< 60\,{\rm degrees}$.



\subsection{Calibration}

As we will review in \S3, our expectation (based on nearly complete
redshift information from the Northern HDF, and photometric redshifts
in the Southern HDF) is that $>90\%$ of spirals in the HDF samples at
$I_{814W} < 23$ mag lie at redshifts $z < 1$. We argue in this section
that this magnitude limit is also approximately that at which bars are
generally detectable in HDF WF/PC2 images on the basis of
signal-to-noise. The importance of these two statements is that local
$B$-band images represent a close rest-frame match to most of the
distant spirals in an $I_{814W}<23$~mag HDF sample\footnote{In fact,
for many of the brighter HDF galaxies studied in the present paper,
local $B$-band calibration data is actually somewhat {\em blueward} of
the observed rest wavelengths for HDF galaxies. However, in local
galaxies, bars are not substantially more visible at rest $V$-band
wavelengths, compared to rest $B$-band. We have confirmed this by
duplicating the present analysis with $R$-band images of galaxies in
the Frei et al. (1996) sample, in which we find a very similar
distribution to that shown in Figure~1. Bar visibility is expected to
drop off markedly blueward of the 4000\AA~ break, and to increases
sharply in the near infrared for most red bars, because of the strong
supression in the visibility of young stellar populations, enhancing
the relative contrast of the generally reddish bars relative to the
disc.}. In other words, our study will be largely unaffected by
``morphological K-corrections'' that play a significant role in the
interpretation of HDF data at $I_{814W}>23$ mag.

The effective $B$-band rest wavelength of most of our sample is 
also convenient because of the public availability of a sample of 
$B$-band CCD data for a ``generic'' sample of bright local galaxies 
(Frei et al. 1996). The Frei sample of local galaxies was observed at a
physical resolution similar to that achieved by HST when probing
high-redshift galaxies.  This sample can therefore be used in order to
investigate the utility of \ba~in discriminating between barred and
unbarred spirals.

Measurements of \ba~for all spiral galaxies in the Frei et al. (1996)
sample are shown in Figure~1.  (In this figure, and throughout the
remainder of this paper we will adopt the terminology of the RC3, and
denote strongly barred spirals as class SB, weakly/tentatively barred
systems as class SAB, and unbarred spirals as class SA.)  It is clear
that the single \ba~parameter is remarkably effective at
distinguishing between strongly barred and unbarred spiral systems.  A
representative cut in \ba~useful for discriminating between strongly
barred and non-barred samples is $(b/a)^2_{\rm bar} = 0.45$.  At all
inclinations SB galaxies have systematically smaller \ba~than SA
systems, although the gulf between the two classes is a fairly strong
function of inclination, and by $i \sim 70\,{\rm degrees}$ the classes
are sufficiently intermingled that our methodology is rendered
ineffective.  As described in \S2.1 this is entirely expected because
of the difficulty in distinguishing highly inclined barred from
unbarred spirals, not only in our objective analysis, but also when
undertaking visual classifications (Naim et al. 1995). SAB systems
typically lie in-between the strongly and weakly barred systems at any
given inclination. In the present paper we adopt an inclination limit
of 60 degrees in defining our sample.  Our methodology does not
distinguish the tentative/weakly barred SAB systems from other
classes, but as described earlier there is good evidence that visual
classification of these systems is particularly subjective and
inherently rather poorly defined in local catalogs (Naim et al. 1995).

\begin{figure}
\centering \epsfig{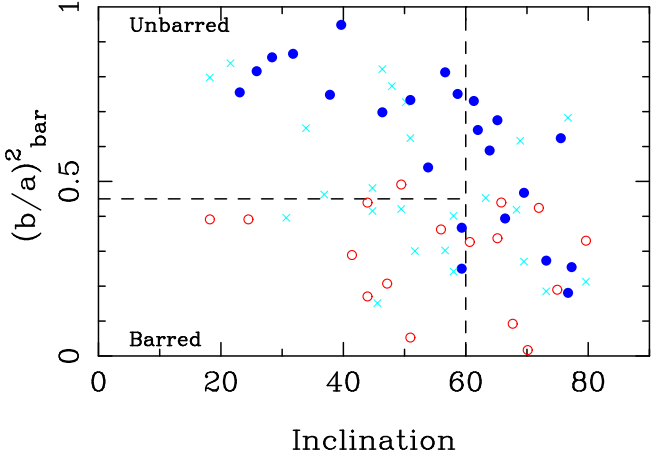}
\caption{\it \ba~plotted as a function of galaxy inclination for the
local spirals in the Frei et al. (1996) B-band CCD imaging
sample. Note how this single parameter is remarkably effective at
distinguishing between barred and unbarred spirals at inclinations
below $\sim 60$ degrees. Both our proposed cut at \ba~ = 0.45 and
inclination cut at 60 degrees are shown. Symbols are keyed to
classifications in RC3. Strongly barred spirals (RC3 class SB) are
shown are open circles, weakly barred spirals (RC3 class SAB) are
shown as crosses, and unbarred spirals (RC3 class SA) are shown as
filled circles.  }
\end{figure}

\section{BARRED GALAXIES IN THE HUBBLE DEEP FIELDS}

We now proceed to discuss the application of the \ba~parameter in the
measurement the bar fraction in the Hubble Deep Fields.  Adopting the
formulation described in Abraham et al. (1996a,b), the Frei et al.
(1996) sample was artificially redshifted within the range $0<z<1.5$
in order to determine the faintest magnitude to which \ba~ can be
determined accurately under the conditions of the HDFs.  The very
clean separation between SA and SB systems in Figure~1 was recovered
to synthetic redshifts associated with $I_{814W}=23.2$ mag, which we
adopted as our magnitude limit. This limit was subsequently confirmed
by an internal analysis of data from the HDFs themselves. In this
procedure, bright spirals in the HDFs were shrunk, faded, convolved
with the PSF, noise-degraded and digitally added to blank portions of
the HDF-North in order to test the magnitude limit at a range of
isophotal sizes spanning the observed size distribution of
$I_{814W}\sim23$ mag spirals.  The robustness of \ba~at our adopted
$I_{814W}$ magnitude limit is shown in Figure~2.

\begin{figure}
\centering \epsfig{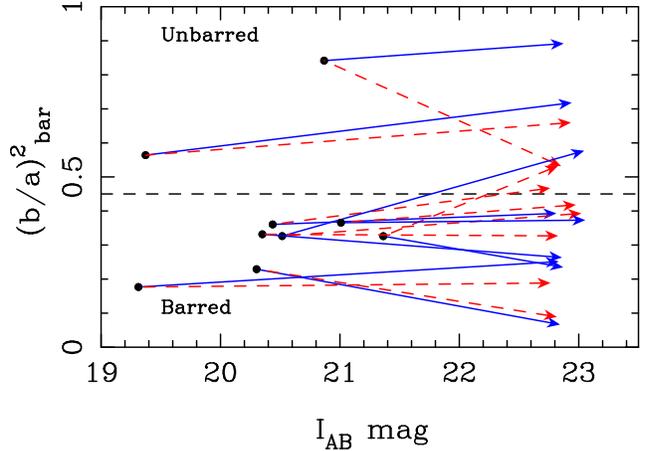}
\caption{\it Tracks showing the recovery of \ba~ measurments for
bright HDF systems degraded to the magnitude limit of our sample. The
predominantly horizontal tracks indicate that, within the range
explored in the present paper, measurements of \ba~ are robust to
signal-to-noise and size degradation. Solid tracks correspond to
galaxies reduced to 60\% of their original size, and dashed tracks to
the galaxies in the extreme case where galaxies are reduced to 30\% of
their original size.  After size reduction and diminution of galaxian
light to our approximate flux limit, galaxies were convolved with the
dithered HST \ib~PSF before being digitally added to a blank portion
of the HDF sky and re-analysed.}
\end{figure}

Prior to measuring \ba, spiral systems were extracted from the galaxy
mix using the Abraham et al (1996a) asymmetry-concentration (A-C)
classifier.\footnote{The morphological region used to extract spirals
in the log A vs. log C classification plane was defined in terms of
linear boundaries, as in Abraham et al (1996a).  Both the early-type
vs. spiral and spiral vs. irr/peculiar/merger lines intersect the
log(A)=0 axis at log(C)=-0.16, with slopes of 9.15 and 2.30,
respectively.} The previously mentioned inclination cut of
$i<60^{\circ}$ was then applied, resulting in a low-inclination sample
of 20 HDF-N and 28 HDF-S spiral galaxies of which 18 (35\%) are
strongly barred (9 in each HDF) according to the $(b/a)^2_{\rm bar}$
discrimator described in Section 2.2. A montage of these galaxies is
presented in Figure 3.

Subsequent visual inspection of these images indicates a possible
uncertainty in the barred identification of, at most, four systems. In
two, features identified as a weak spiral arm emanating from a bar
could conceivably be a companion or optically superposed galaxy
aligned at right angles to the main body of the galaxy, leading to a
spurious classification. In one case, a strong bar is clearly seen,
but it is not obviously accompanied by spiral structure in the outer
regions, making classification as a peculiar possibly appropriate.  In
the final case, the inner isophotal structure could be described as
perhaps more lens-like than bar-like. (Similar systems are sometimes
categorized as barred in local catalogues -- the distinction between
nuclear bars and lenses can be rather subjective). Each of these
uncertainties would diminish the proportion of barred spirals at
$z\simeq$0.5 and not affect the overall conclusion of the paper.

\begin{figure*}
\centering
\epsfig{figure=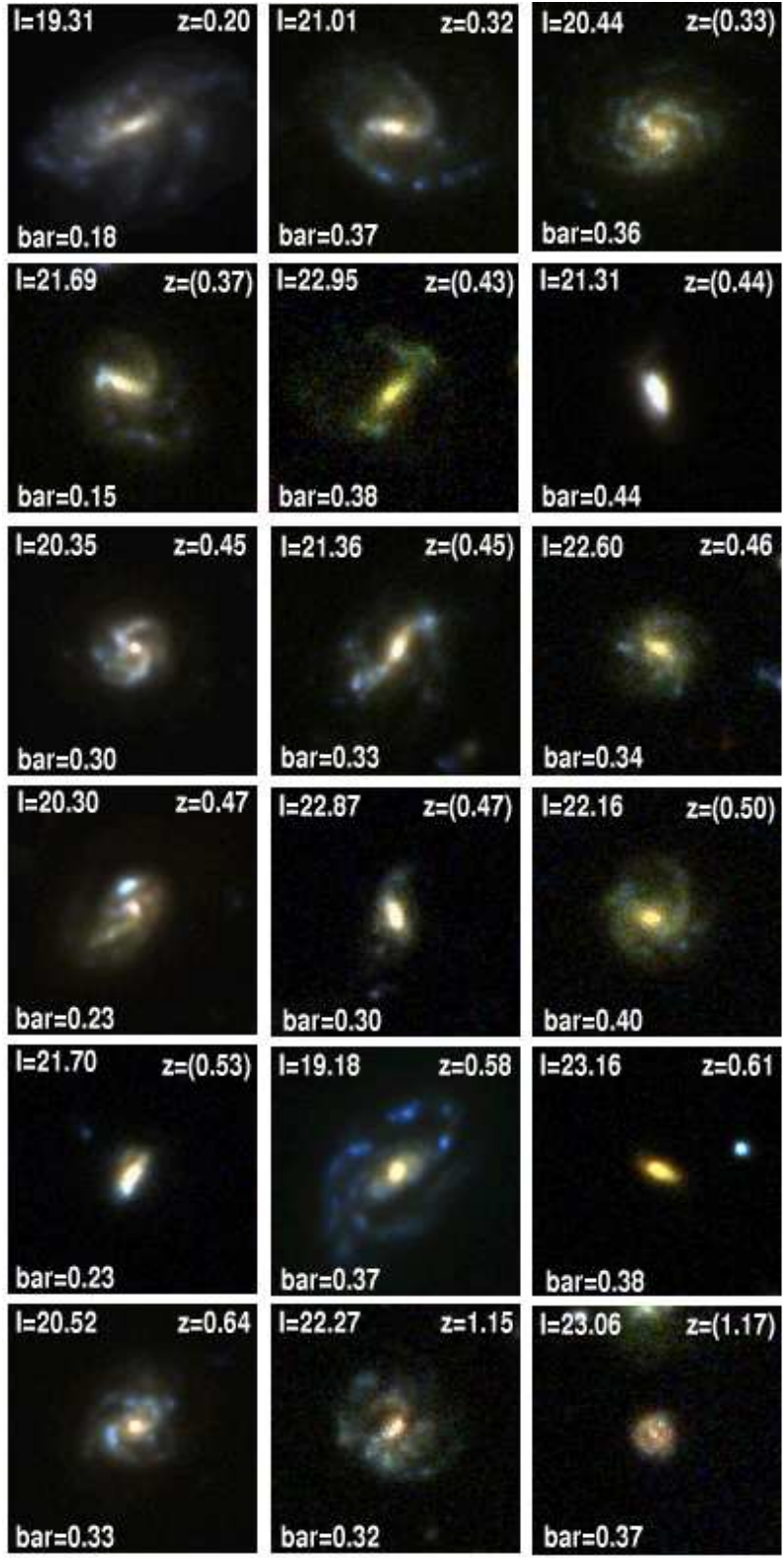,width=21cm,angle=-90} 
\caption{\it All spiral galaxies with $I_{814W}<$23.2, inclination
$i<60^{\circ}$ and $(b/a)^2_{\rm bar}<0.45$ in the northern and
southern Hubble Deep Fields ordered according to redshift. True colour
images are shown, constructed from the $B_{450W}$, $V_{814W}$, and
$I_{814W}$-band data. Each panel represents a field of 5.1 arcsec on a
side. The redshift and value of $(b/a)^2_{\rm bar}$ is indicated
(photometric redshifts are in parenthesis).}
\end{figure*}

A more important question is whether, in selecting the face-on spiral
sample using the A-C estimator, some barred objects have been
missed. This would be immaterial in estimating the fraction of
barred/unbarred galaxies in the HDF unless, by some means, barred
systems were preferentially excluded in extracting the A-C spiral
sample. We have investigated this by visually classifying all galaxies
in the HDF-S/N to $I_{814}$=23.2 mag. As in Abraham et al (1996a,b),
the agreement between the A-C spiral selection and the visual spiral
selection was excellent. Visual classification would add two barred
spirals to the montage shown in Figure~3, and as both systems are
bright and at low redshift, inclusion of these objects would actually
slightly strengthen the conclusions of this paper.

As described in the Introduction, because of disagreements with regard
to the definitions and proportions of weakly-barred spirals in local
catalogues, one cannot robustly map the $(b/a)^2_{\rm bar}$
discriminator onto the SA/SAB/SB scheme to compare with the local
barred spiral fraction. Therefore a more productive way forward is a
{\em self-consistent internal comparison of the redshift distribution
of barred and unbarred spirals within the HDF datasets.}  In order to
construct this for our magnitude-limited sample, we must augment the
spectroscopic redshifts available for the HDFs with photometric
redshift estimates, particularly for HDF-S for which published
spectroscopic data is limited. In the HDF-N all but two systems in our
sample have known spectroscopic redshifts, while in the HDF-S the
situation is almost exactly reversed, with only two galaxies in our
sample having currently known spectroscopic redshifts (Glazebrook et
al. 1998, in preparation). As barred spirals are indistinguishable
from their non-barred counterparts on the basis of colour (de
Vaucouluers 1961) we do not expect errors in the determination of
photometric redshifts to significantly bias our results.

In the case of the HDF-N, we used the compilation of Wang et al (1998)
to provide photometric redshifts for the 2 remaining HDF-N face-on
spirals to $I_{814}$=23.2 without spectroscopic data.  For the HDF-S
sample we have computed our own photometric redshifts, based on a
template fitting method (Brinchmann et al 1998, in preparation)
similar to that used by Fern{\'a}ndez-Soto et al (1998). We have
compared our estimates in the HDF-N with those obtained using the
linear fitting formulae from Wang et al (1998).  Our determinations
agree well with a negligible mean offset and an RMS scatter of only
0.1 in redshift, which is adequate for our purposes. Both methods have
been extensively tested against spectroscopic data in the HDF-N (Wang
et al 1998) and within the redshift range concerned agree remarkably
well.
 
\begin{figure*}
\centering
\epsfig{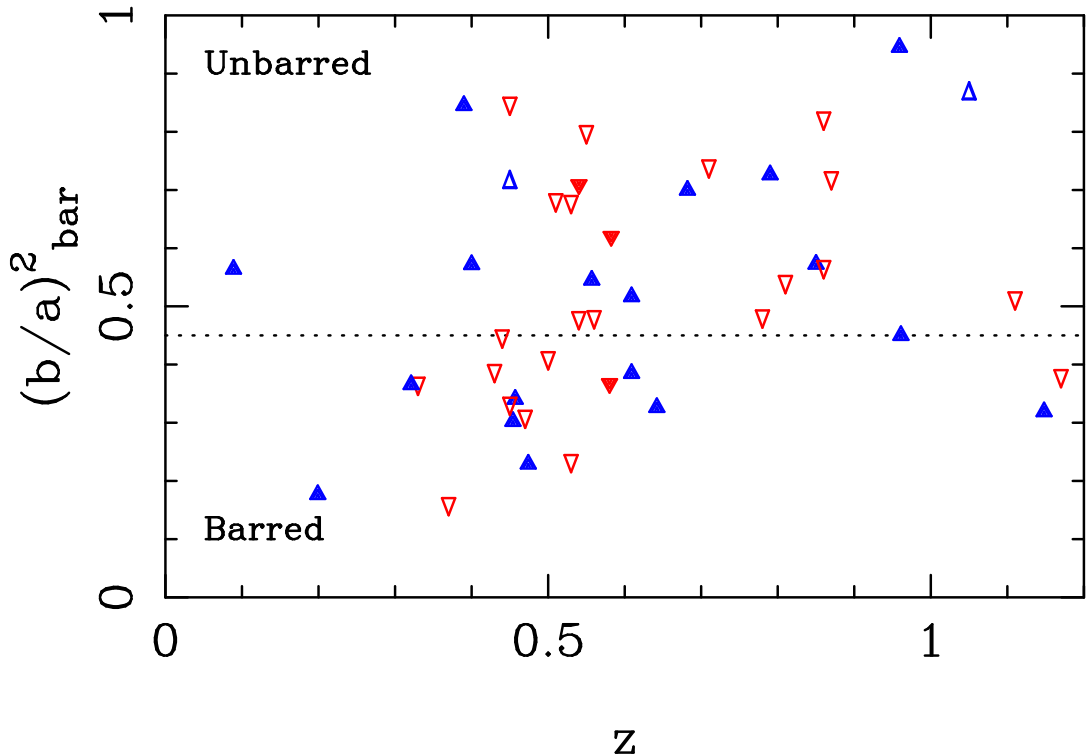}
\caption{\it The bar strength estimator $(b/a)^2_{\rm bar}$ plotted as
a function of spectroscopic redshift (filled symbols) or photometric
redshift (open symbols) in the northern (triangles pointing upward)
and southern (triangles pointing downward) Hubble Deep Fields. Note
the marked gradient in the proportion of barred systems with redshift,
beginning at $z\sim 0.5-0.6$.}
\end{figure*}

Figure 4 shows the redshift distribution of barred and unbarred
spirals in the Hubble Deep Fields North and South. {\em This diagram
reveals a striking decline in the proportion of barred examples beyond
a redshift $z \sim 0.5$.} This cannot be due to uncertainties in using
photometric redshifts for the HDF-S as the same effect is seen in both
HDF samples and the HDF-N is spectroscopically complete at the 90\%
level. Formally, the redshift distributions of the barred and unbarred
samples selected on the basis of \ba~ are inconsistent at the 99.8\%
confidence level from a Kolmogorov-Smirnov test. We conclude that
Figure~4 shows strong evidence for differential evolution in the
abundance of barred spirals at high redshift, in the sense of a marked
decrease in the proportion of barred spirals beyond $z\simeq$0.5.

\section{DISCUSSION}

As mentioned in the Introduction, there are two basic criteria that
must be met for the bar-forming instability to be effective: the
material in the disk must be self-gravitating, and it must be
following relatively well-ordered orbits, without excessive random
motion.  Thus, one possible explanation for the deficit of barred
galaxies beyond $z \sim 0.5$ is that disks at these redshifts have not
yet accreted sufficient material to be self-gravitating.
Unfortunately, it is difficult to measure the degree to which the
masses of disks dominate even in nearby galaxies with plentiful
kinematic data (e.g.\ Casertano \& van Albada 1990, Freeman 1993, van
der Kruit 1995), so it will be very hard to test this hypothesis
observationally on these high-redshift objects.

The second possibility is that the bar instability is suppressed due
to larger random motions in the material making up the high-redshift
disks (e.g. Ostriker \& Peebles 1973).  If a disk has only recently
formed, one might expect the orbits of the material in it to reflect
the somewhat stochastic process by which it was assembled, leading to
relatively large random components in their motion.  Such a ``hot''
disk would prove resistant to the bar instability, explaining the
deficit of barred galaxies at higher redshifts.

A final possibility is that the mechanisms that destroy bars are more
efficient at high redshift.  For example, Pfenniger (1991) has shown
that a merger with a compact companion can heat a disk to a point
where any bar is destroyed; a more major merger, on the other hand,
would entirely destroy the disk, leaving an elliptical galaxy
(e.g. Barnes 1992).  In any hierarchical picture of galaxy formation,
mergers are more common at high redshift, and generally consist of
collisions with less-developed small galaxies (Navarro, Frenk \& White
1994).  We would therefore expect a higher rate of conversion of
barred galaxies into unbarred disk galaxies at high redshift than we
see today.

The efficient suppression of a high-redshift bar may not even require
the intervention of an outside agency such as a merger.  As discussed
in the Introduction, bars are known to be unstable entities, which
evolve into spheroidal bulge-like structures.  This spontaneous
buckling process introduces a random element into the motions of stars
similar to that induced by a minor merger, and so acts to suppress
subsequent bar formation. In an internal-colour based analysis of
bright HDF-N galaxies (Abraham et al. 1999), it was concluded that
high-redshift bulges are the oldest components of the galaxies at the
epoch of observation.  If these bulges formed from an early generation
of bars at a redshift higher than that probed in the present study,
then further bar formation will be inhibited until sufficient new
material has been added to the disk on well-ordered orbits to allow
the instability to act.  Thus, the deficit of barred systems at $z >
0.5$ may represent a temporary lull in the phenomenon while the
galaxies recover from their first bout of bar formation.

Similarly, Hasan \& Norman (1990) have shown that the presence of a
central mass of a few percent of the total disk mass will suppress the
bar-forming instability.  In addition, a bar provides an effective
mechanism for channeling gas toward the centre of a galaxy (e.g.\
Friedli \& Benz 1993).  Thus, we might expect a bar to eventually
destroy itself by feeding sufficient material toward the centre of its
galaxy to suppress the instability.  Sellwood \& Moore (1998) have
explored this scenario in some detail, and have pointed out that
several mechanisms exist by which a second generation bar could be
produced, perhaps explaining the prevalence of bars at low redshifts.

\section{CONCLUSIONS}

We have defined a simple measure of bar strength suitable for probing
the internal structure of faint galaxies. Our technique is based on
measuring parameters sensitive to the photometric signatures of barred
spiral structure. These observables can be combined to yield the {\em
physical} axial ratio of a bar, under the assumption of an elliptical
bar embedded within a round, thin disk. Where these assumptions are
not a good approximation, the estimator still yields a perfectly
quantitative, objective parameter that appears to closely track visual
estimates of bar strength.

Our parametric bar estimator, \ba, has been tested against local
samples of spirals, and against artificially redshifted spirals under
the conditions of the Hubble Deep Fields. For reasonably
low-inclination systems ($< 60$ degrees), spirals classed locally as
SA are cleanly-separated from systems classed as SB. We demonstrate
that distinguishing between barred and unbarred spirals on the basis
of \ba~is possible to $I=23.2$ in the Hubble Deep Fields.

The release of the Southern Hubble Deep Field has increased the sample
of $I<23$ mag low-inclination galaxies substantially, allowing the
first detailed investigation of redshift evolution in the barred
spiral fraction.  By combining measurements of \ba~ with spectroscopic
and photometric redshifts in the Hubble Deep Fields, a striking
decrease in the proportion of barred spirals as a function of redshift
beyond $z=0.5$ has been discovered. Our result supports the
observation by van den Bergh et al. (1996) that barred spirals are
deficient at faint magnitudes in the northern HDF. 

The physical mechanisms responsible for the absence of barred spirals
at high redshifts is unclear.  Possibilities we have discussed include
dynamically hotter (or increasingly dark-matter dominated)
high-redshift discs, and an enhanced efficiency in bar destruction at
high redshifts.  For all these scenarios, the present result provides
a clear observational benchmark.  Our investigation has brought
forward the epoch at which the conventional Hubble system is observed
to be in place, from $z \sim 1$ (based on the the abundance of
morphologically peculiar systems in the LDSS/CFRS redshift survey
described in Brinchmann et al. 1988) to $z \sim 0.5$.

{\bf Acknowledgments} We thank Bob Williams and Harry Ferguson at
STScI for their encouragement, and the entire STScI for their heroic
efforts in releasing the HDF-S data promptly as advertised. We also
thank Alar Toomre, Len Cowie, Malcolm Longair and Sidney van den Bergh
for useful discussions. RGA and MM acknowledge PPARC for support under
the Advanced Fellowship programme. JB acknowledges receipt of an Isaac
Newton Studentship.

\end{document}